\documentstyle[pre,aps,preprint,epsfig]{revtex}       
\tightenlines  

\newcounter{Figure}

\def\gg{g^{(2)}(1,2,\rho)}

\def\ggtwoone{g^{(2)}(1,2,\rho_1,\rho_1)}
\def\ggtwotwo{g^{(2)}(1,2,\rho_1,\rho_2)}
\def\gone{g^{(2)}(1,2,\rho_1)}

\def\rr{\rho^{(2)}(1,2,\rho)}
\def\rrtwo{\rho^{(2)}(1,2;\rho_1,\rho)}
\def\rrtwoone{\rho^{(2)}(1,2;\rho_1,\rho_1)}
\def\rrtwotwo{\rho^{(2)}(1,2;\rho_1,\rho_2)}
\def\rrthree{\rho^{(2)}(1,2;\rho_1,\rho_2,\rho)}
\def\rrthreetwo{\rho^{(2)}(1,2;\rho_1,\rho_2,\rho_2)}

\def\ftwo{\Phi(\rho_1,\rho)}
\def\fthree{\Phi(\rho_1,\rho_2,\rho)}

\newcounter{Table}

\begin{document}
\title{Enhanced Saturation Coverages in Adsorption-Desorption Processes}
\author{Paul R. Van Tassel$^1$, Pascal Viot$^2$,  Gilles Tarjus$^2$,\\
Jeremy J. Ramsden$^3$ and Julian Talbot$^4$\\
{$^1$\it Department of Chemical Engineering and Materials Science,}\\ 
{\it Wayne State University, Detroit, MI 48202}\\
{$^2$\it Laboratoire de Physique Th{\'e}orique des Liquides}\\
 {\it Universit{\'e} Pierre et Marie Curie}\\
{\it 4, place Jussieu  75252 Paris Cedex 05  France}\\
{$^3$\it Department of Biophysical Chemistry, Biozentrum, CH-4056
Basle, Switzerland}\\
{$^4$\it Department of Chemistry and Biochemistry,}\\
{\it Duquesne University, Pittsburgh, PA 15282-1503}\\[1.0cm] }
\maketitle
\begin{abstract}
Many  experimental  studies of protein    deposition on solid surfaces
involve  alternating  adsorption/desorption steps. In   this paper, we
investigate the effect of a desorption step (separating two adsorption
steps)  on  the   kinetics, the  adsorbed-layer    structure,  and the
saturation density.  Our    theoretical approach involves a    density
expansion of  the pair distribution  function and an application of an
interpolation formula to estimate the saturation density as a function
of the density at which  the desorption process commences, $\rho_1$, and
the density of the depleted configuration, $\rho_2$. The theory predicts
an  enhancement  of the  saturation density  compared  with  that of a
simple, uninterrupted RSA  process   and a maximum in  the  saturation
density when $\rho_2=\frac{2}{3}\rho_1$.  The  theoretical results  are in
qualitative  and in semi-quantitative   agreement with the results  of
numerical simulations.
\end{abstract}

\newpage
\section{Introduction}
The  adsorption of  proteins  and  colloids from  solution onto  solid
surfaces often  involves  strong departure from   equilibrium behavior
\cite{andrade86}. Many experimental studies have attempted to quantify
the kinetics \cite{wahlgren95,buijs96,ramsden93a,robeson96}   of   the
adsorption  process,   as  well  as   the structure  of  the  adsorbed
configurations  \cite{adamczyk94,cullen94}.    Protein  adsorption has
also   been the   subject  of a  number of    theoretical studies.  In
particular,  the  Random      Sequential   Adsorption (RSA)      model
\cite{evans93} is able   to  account for  a  number of  experimentally
observed properties of protein adsorption, such as irreversibility and
non-linear surface  blockage \cite{ramsden93, ramsden93a, adamczyk94}.
Quantitative expressions are now available for  the kinetics of RSA of
spherical \cite{schaaf89,tarjus91,jin94} and  non-spherical  particles
\cite{ricci92,adamczyk96} on  planar interfaces and these  models have
been shown  to provide an  excellent description of the  adsorption of
proteins            \cite{ramsden93,ramsden93a,jin94},        colloids
\cite{adamczyk96,johnson95} and  other particles.  More  recently, the
RSA  models  have  been  extended to  allow   for  the  possibility of
post-adsorption conformational change
\cite{vantassel94, vantassel96}.

It has been observed experimentally that a fraction of the proteins in
an adsorbed layer may be removed by rinsing with  a buffer solution or
by eluting with a surfactant solution
\cite{kurrat94,wannerberger96,wahlgren95,buijs96,malmsten96,wahlgren98}.
The   influence of a desorption step   on subsequent adsorption is not
well   understood.  In particular,  one would  like   to  know how the
structure of the adsorbed layer  is affected by the desorption process
and how  this  influences the  kinetics when adsorption  resumes.  One
anticipates, due to the non-equilibrium nature of the adsorption, that
the intermediate desorption  step may result  in behavior that differs
non-trivially  from an uninterrupted process.  It  is the goal of this
work to  investigate this by considering a  multistep RSA process that
begins with the irreversible adsorption of  spherical particles onto a
planar  surface up  to a density  $\rho_1$.  Adsorbed particles are then
removed randomly until the density  falls to $\rho_2$.  We then initiate
a second  irreversible adsorption process  beginning from the depleted
configuration and investigate how the final saturation density depends
on $\rho_1$ and $\rho_2$. The memory effect of the RSA process is expected
to  result in a   final density that  is  different  from  that of  an
uninterrupted RSA process.

In the theoretical  part of the study,  we use a distribution function
approach to  determine the   influence  of the   multistep  adsorption
process  on the structure  of the adsorbed  particle configuration.  A
density expansion of  the  pair  distribution  function leads  to   an
approximate description of the structure up  to the maximum density in
the  initial  adsorption step. We then  show  that if particles desorb
randomly  from  the adsorbed  configuration,  the  radial distribution
function is unchanged.  We use this  result to estimate  the available
surface function of the configurations  generated by the re-adsorption
process and develop approximate interpolant  formulas that allow us to
estimate  the final  saturation density.  The theory  is then compared
with   numerical   simulations  of  the   process  in   one  and   two
dimensions. Finally, we use  these expressions to determine the extent
of desorption  needed to optimize the  time required to  reach a given
adsorbed density.

\section{Theory}

We  consider a three-step adsorption/desorption/readsorption  process.
The initial phase consists of  the irreversible addition of  spherical
particles of diameter $\sigma$ on an initially empty planar interface to a
density  $\rho_1 < 0.547$  (in this  work we  will  use $\rho$ to denote a
dimensionless density, $\rho = \frac{\pi}{4}\sigma^2 N  / A$ where $N$ is the
number  of particles adsorbed  in an area  $A$). During this phase the
particles are  placed according to  the usual RSA  rules i.e., a trial
position is selected from a  uniform random distribution. If the trial
particle  does not overlap with any  previously placed particles it is
accepted and remains permanently fixed.  Otherwise it is rejected.  In
the subsequent    desorption   step,   particles   in   the   adsorbed
configuration are removed   at random  until   the final  density   is
$\rho_2$.  In  the final step,  the  irreversible  adsorption process is
resumed.  New particles  can  be added  to the depleted  configuration
until the jamming limit is reached. As a result of the inherent memory
of   the RSA   process,      the coverage  in   this     final  state,
$\rho_{\infty}(\rho_1,\rho_2)$ is, in general, different  from the usual jamming
limit for  RSA that occurs  without the intermediate  desorption step,
namely 0.547
\cite{evans93}.

The adsorbed configurations are characterized by a set of distribution
functions $\rho_n({\bf r}_1,{\bf r}_2,\cdots,{\bf r}_n;\rho)$, which represent
the density of finding  $n$ unspecified particles among $N$  particles
defining  the system,  at  the positions ${\bf  r}_1,{\bf r}_2,\cdots,{\bf
r}_n$ within  an   area $A$  such   that $\rho=N/A$.  Centered   on each
pre-adsorbed particle is a circular  exclusion region of diameter $\sigma$
into  which it is  impossible to  insert the  center of  an additional
particle.   The probability of finding   a   cavity of diameter  $2\sigma$
centered on ${\bf r}_1$ that is free  from the centers of pre-adsorbed
particles is related to the distribution functions via
\begin{equation}
\Phi({\bf r}_1^0) = 
1 + \rho\int d2\;f_{12} + \frac{1}{2!}\int d2\;d3 f_{12}f_{13}
\rho^{(2)}(2,3,\rho) + \frac{1}{3!}\int d2\;d3\;d4 f_{12}f_{13}f_{14}
\rho^{(3)}(2,3,4,\rho)+\cdots
\label{phi}
\end{equation}
where  $d1$    is  shorthand for   $d{\bf   r}_1$  etc., and $f_{ij}\equiv
f(r_{ij})=\exp(-u(r_{ij})/kT) -  1$  is the  Mayer  function. When the
interparticle  potential,  $u(r_{ij})$, is  taken for  non-overlapping
hard spheres of diameter $\sigma$, the Mayer function becomes:
\begin{equation}
f(r_{ij})= \left\{ \begin{array}{c}
-1,\;\; r_{ij} < \sigma \\
0,\;\; r_{ij} > \sigma
                   \end{array}
           \right.
\end{equation}
Equation \ref{phi} applies to any configuration, equilibrium 
or not, of spherical particles.

Our  strategy  is     to determine  the    relationship  between   the
configurations before and after the desorption process. This allows us
to find   an approximate  expression   for $\Phi$  which we  can  use to
estimate the saturation coverage as a function of $\rho_1$ and $\rho_2$.

\subsection{Initial Adsorption}

The kinetics during the irreversible adsorption phase
are given by
\begin{equation}
\frac{d\rho}{dt} = k_a \Phi
\label{rate1}
\end{equation}
where $k_a$ is the adsorption rate constant.

To   solve for the    kinetics, one writes down  a   hierarchy of rate
equations for the distribution functions \cite{schaaf89,tarjus91}.  In
particular, the pair density evolves according to
\begin{equation}
\frac{\partial\rr}{\partial t} = 2 k_a \Phi(1,2^o)
\label{rate2}
\end{equation}
where $\Phi(1,2^o)$ is the  probability density function associated with
finding an  unspecified  particle  at ${\bf  r}_1$  and   a cavity  of
diameter   $2\sigma$ free from  particle   centers  at ${\bf r}_2$.   This
function may  also be expressed  in terms  of Mayer  functions and the
n-particle density functions:
\begin{equation}
\Phi(1,2^o) =  (1+f_{12})\bigl[\rho
+\int d3\; f_{23}\rho^{(2)}(1,3,\rho) +O(\rho^3)\bigr]
\end{equation}
Substituting (\ref{rate1}) in (\ref{rate2}) 
to remove the time dependence and noting that $\rho^{(2)}(1,3,\rho)=\rho^2(1+f_{13})(1+O(\rho^2))$ yields
\begin{eqnarray}
\frac{\partial\rr}{\partial\rho}
& = & \frac{\Phi(1,2^o)}{\Phi(1^o)}\nonumber\\
& = & \frac{2(1+f_{12})\bigl[\rho
+\rho^2\int d3\; f_{23}(1+f_{13})+O(\rho^3)\bigr]}{1+\rho\int d2\; f_{12}
+O(\rho^2)}\nonumber\\
& = & 2(1+f_{12})\bigl[\rho+\rho^2\int d3\;f_{13}f_{23}
+ O(\rho^4)\bigr]
\label{r2r}
\end{eqnarray}

Integrating between 0 and $\rho$ gives the pair density function to
third order in $\rho^3$: 
\begin{equation}
\rr = (1+f_{12})\bigl[\rho^2+\frac{2}{3}\rho^3\int d3\;f_{13}f_{23}
+ O(\rho^4)\bigr]
\label{rsa2}
\end{equation}

\subsection{Desorption}

The subsequent 
desorption process results in 
the removal of a fraction of the 
particles from an initial configuration at a density $\rho_1$ 
so that the final density is $\rho_2$. During this step, 
the density evolves according to
\begin{equation}
\frac{d\rho}{dt} = -k_d \rho
\end{equation}
where $k_d$ is the desorption constant and, if no other relaxation
processes (such as surface diffusion) are operative during the desorption,
the pair density function evolves according to
\begin{equation}
\frac{\partial{\rrtwo }}{\partial t}  = -2 k_d \rrtwo
\end{equation}
where we indicate explicitly the dependence of the pair density on the
density prior to   desorption,  $\rho_1$.  It  follows  from   these two
equations that
\begin{equation}
\frac{\partial{\rrtwo }}{\partial\rho}  = \frac{2}{\rho} \rrtwo
\end{equation}
Integrating between the  initial ($\rho_1$) and final densities ($\rho_2$)
yields
\begin{equation}
\frac{\rrtwotwo }{\rrtwoone } = \frac{\rho_2^2}{\rho_1^2}
\label{gequiv}
\end{equation}
or
\begin{equation}
\gone = \ggtwoone = \ggtwotwo
\end{equation}
where $\rr = \rho^2 \gg $ defines the radial distribution function.  
This physically intuitive result 
implies that the structure of an arbitrary configuration 
is not altered by desorption, as long as the particles desorb
randomly. Further consequences of this result and its connection
with quenched annealed systems are discussed elsewhere \cite{vantassel97}.

\subsection{Re-Adsorption}

We now wish to relate the properties of configurations during
the re-adsorption step to those of the configurations in the
initial adsorption step. 
Up to third order in density, the differences between the two configurations
come from the term $\rr$.

Integrating (\ref{r2r}) between $\rho_2$ and $\rho>\rho_2$,
\begin{equation}
\rrthree = \rrthreetwo + (1+f_{12})(\rho^2+\rho_2^2)
+(1+f_{12})(\frac{2}{3}\rho^3 - \frac{2}{3}\rho_2^3)\int d3\;f_{23}f_{13}
+O(\rho^4)
\end{equation}

Using (\ref{gequiv}) to relate the pre- and post-desorption
pair distribution functions, we find
\begin{equation}
\rrthree = (1+f_{12})\bigl[\rho^2+\frac{2}{3}[\rho^3+\rho_2^2(\rho_1-\rho_2)]
\int d3\;f_{13}f_{23} + O(\rho^4)\bigr]
\end{equation}
Substituting in (\ref{phi}) then yields
\begin{eqnarray}
\fthree & =& 1 + \rho\int d2 f_{12} + \frac{1}{2} \rho^2 \int d2\;d3\; (1+f_{23})\\
& & +\frac{1}{3}[\rho^3+\rho_2^2(\rho_1-\rho_2)]\int d2\;d3\;d4\; 
f_{12}f_{13}f_{24}f_{34}(1+f_{23}) \\ 
& &  + \frac{1}{6}\rho^3\int d2\;d3\;d4\; f_{12}f_{13}f_{14}
(1+f_{23})(1+f_{24})(1+f_{34}) + O(\rho^4)\\
& = & \Phi^{\rm RSA}(\rho) + A_{32}\rho_2^2(\rho_1-\rho_2), \;\; 
\rho\geq\rho_2
\label{phiapp}
\end{eqnarray}
where
\begin{equation}
A_{32} = \frac{128}{3\pi^2}\biggl(\frac{\pi\sqrt 3}{2} - \frac{9}{4}\biggr)
\end{equation}
is the portion of the 3rd order RSA coefficient due to first order
corrections of $g^{(2)}$ \cite{schaaf89}. We note that $\fthree$ 
correctly reduces to 
$\Phi^{\rm RSA}(\rho)$ in both the limit where $\rho_1 = \rho_2$
(no desorption) and $\rho_2=0$ (total desorption). 

During desorption, it is easy to show that
\begin{equation}
\ftwo = \Phi^{\rm RSA}(\rho) + A_{32}\rho^2(\rho_1-\rho)+O(\rho^3), \;\; 
\rho_1\geq\rho\geq\rho_2
\end{equation}
We plot that $\Phi$ during adsorption, $\Phi^{\rm RSA}$, desorption $\ftwo$, 
and re-adsorption, $\fthree$, in Fig. \ref{fig:avsf}. One can also show that
\begin{equation}
\frac{d\Phi}{d\rho}\biggl|_{1,{\rm ads}} = 
\frac{d\Phi}{d\rho}\biggl|_{1,{\rm des}} + 
A_{32}\rho_1^2
\end{equation}
where the subscripts ads and des indicate the adsorption and desorption
branches, repsectively and since $A_{32}>0$, we have
\begin{equation}
\frac{d\Phi}{d\rho}\biggl|_{1,{\rm ads}} > 
\frac{d\Phi}{d\rho}\biggl|_{2,{\rm des}}
\end{equation}
so the slope of 
$\Phi$ during desorption is more negative.
One has, furthermore, that
\begin{equation}
\frac{d\Phi}{d\rho}\biggl|_{2,{\rm ads}} = \frac{d\Phi}{d\rho}\biggl|_{2,{\rm des}}
- A_{32}\rho_2(2\rho_1-3\rho_2)
\end{equation}
so
\begin{equation}
\frac{d\Phi}{d\rho}\biggl|_{2,{\rm ads}} > \frac{d\Phi}{d\rho}\biggl|_{2,{\rm des}},
\;\; \rho_2 < \frac{2}{3}\rho_1
\end{equation}
\begin{equation}
\frac{d\Phi}{d\rho}\biggl|_{2,{\rm ads}} < \frac{d\Phi}{d\rho}\biggl|_{2,{\rm des}},
\;\; \rho_2 > \frac{2}{3}\rho_1
\end{equation}
and thus the change in $\Phi$ may be more pronounced in either desorption
or re-adsorption.
It is clear that in either case, the available surface function is enhanced
compared with the simple RSA value (at least to third order in density) 
since
\begin{equation}
\Phi(\rho_1,\rho_2,\rho)-\Phi^{\rm RSA}(\rho)=
A_{32}\rho_2^2(\rho_1-\rho_2) > 0, \;\;\; \rho>\rho_2
\end{equation}

To obtain quantitative results, we consider the interpolation formula
\begin{equation}
\fthree = [1+A_{32}\rho_2^2(\rho_1-\rho_2)](1-x)^3(1+a_1 x + a_2 x^2)
\label{interpolant}
\end{equation}
where
\begin{equation}
x = \rho/\rho_{\infty}
\end{equation}
This form is
consistent with the asymptotic kinetics,
\begin{equation}
\rho_{\infty}-\rho \sim t^{-1/2}
\label{asymp}
\end{equation}
Finally, (\ref{interpolant}) correctly reduces to the equation used to
determine $\rho_{\infty}$ for simple RSA when  $\rho_1=\rho_2$ or when $\rho_2=0$.
Expanding the   interpolant  in  powers   of $\rho$ and    equating  the
coefficients  with  those of   (\ref{phi})   yields an   estimate  for
$\rho_{\infty}( \rho_1,\rho_2)$. As can be seen in Figure
\ref{fig:jlden}, the saturation coverage is a monotonically increasing
function of $\rho_1$ and exhibits a maximum versus $\rho_2$. Actually,
the saturation density increases monotonically with the combination
$\rho_2^2(\rho_1-\rho_2)$, so a maximum occurs for $\rho_2 =
(2/3)\rho_1$.

In one dimension, the coefficient $A_{32}$ takes the value 2/9. 
Unfortunately, however,
the analogous interpolation formula, $\Phi(\rho) = 
[1+A_{32}\rho_2^2(\rho_1-\rho_2)](1-x)^2(1+a_1 x + a_2 x^2)$ 
does not yield physically acceptable
estimates of the jamming density, even in the simple RSA case.

\section{Simulation}

The adsorption-desorption process was simulated using techniques
that have been described previously. Disks are deposited randomly and
uniformly in a periodic cell of area $A$, subject to the usual RSA constraints of
no overlap and no desorption, until the coverage
reaches the preset value, $\rho_1$. In the desorption step,  
randomly selected 
particles are removed until the coverage falls to $\rho_2$. The
RSA process is then restarted and continued for a total of $N_{\rm trial}$
attempts. In this way we obtain the time-dependent coverage, $\rho(t)$,
where $t = a_{\rm rel}i$, $a_{\rm rel}=\pi\sigma^2/4A$ 
and $i$ is the number of attempts that have
been made. This function is then averaged over a number, $N_{\rm av}$,
of independent runs. The coverage in the jamming limit was estimated by
assuming that at long times the kinetics are described by the usual
asymptotic kinetics, (\ref{asymp}).

The simulation procedure in one-dimension is somewhat different. 
Rods of unit length are deposited randomly on a line of length $L$
subject to the usual constraints of no overlap and no desorption. When
the jamming limit coverage is nearly reached, all the remaining available
gaps, each of which can by now accommodate at most one additional rod,
are filled by inserting rods randomly and uniformly within them. 
Partial desorption is then simulated by looping through all the
adsorbed particles and removing each with a probability $p_{des}$. 
This is achieved by generating a uniform random number between zero
and one. If this number is less than $p_{des}$ the particle is
removed. The depleted configuration then serves as the initial
configuration for a new irreversible adsorption process which
continues until jamming. If $N$ denotes the number of adsorbed
rods, then the coverage is computed from $\rho = N/(L-1)$. Since 
the jamming limit can be reached exactly in the one-dimensional 
system, the final results are expected to be more accurate than
in two dimensions.

\section{Results}

In both one and two dimensions, the new jammed state has a higher
coverage than that of simple RSA, although the effect is smaller
than predicted by the theory. In one dimension, Figure \ref{fig:sim1d} 
shows that the maximum density
(just over 0.78) occurs for $p_{des}\approx 0.4$ or $\rho_2\approx 0.6$,
which is consistent with, although a little larger than, 
the theoretical prediction of $\rho_2=(2/3)
\rho_1$.

In two dimensions (Figure \ref{fig:sim2d}), 
the enhancement is not statistically significant
for $\rho_1=0.4$. For $\rho_1=0.53$ the effect is clearly evident
and the maximum final coverage occurs for $\rho_2\approx 0.4$ while
the theoretical estimate is 0.353.

That the theory overestimates the enhancement is not totally surprising.
The estimate of the saturation coverage provided by the interpolation
formula for simple RSA (0.553) exceeds the actual value (0.547). 
Moreover, the
pair density function is taken only to second order in $\rho$.

\section{Application}

An interesting consequence of the findings of this article is that
it may be possible to achieve a given surface coverage 
using an intervening desorption process in a shorter time than
with a continuous adsorption process. We can readily quantify this
idea by minimizing the 
time required to reach a given surface coverage in an 
adsorption-desorption-adsorption process by optimizing 
 densities at which the desorption step begins,
$\rho_1$, and ends, $\rho_{2}$. 
Let $t_f$ denote the time at which the adsorbed density is $\rho_f$. In a
process with a single desorption step
\begin{eqnarray}
t_f & = & t_{\rm ads} + t_{\rm des} + t_{\rm reads}\\
&=&\int_0^{\rho_1}[\Phi^{RSA}(\rho)]^{-1}d\rho + \int_{\rho_1}^{\rho_{2}}
[k_d\rho]^{-1}d\rho + 
\int_{\rho_{2}}^{\rho_f}[\Phi(\rho_1,\rho_{2},\rho)]^{-1}d\rho
\end{eqnarray}
We find the minimum time to reach the density $\rho_f$ by 
by optimizing the 
density at the beginning of the desorption step and that after
desorption. As we will demonstrate, the amount adsorbed in such a
process can exceed that in a continuous adsorption process without 
interruption. 
The required densities are found from a numerical solution of the
equations
\begin{eqnarray}
\frac{\partial t_f}{\partial\rho_1} = 0 \\
\frac{\partial t_f}{\partial\rho_{2}}=0
\end{eqnarray}
This is achieved by iterating the following matrix equation:

\[\left[\begin{array}{c}
\rho_1\\
\rho_{2}
\end{array}
\right]^{\rm new}=\left[\begin{array}{c}
\rho_1\\
\rho_{2}
\end{array}
\right]-\left[\begin{array}{cc}
\frac{\partial^2 t_f}{\partial\rho_1^2} & 
\frac{\partial^2 t_f}{\partial \rho_1\partial\rho_{2}} \\
\frac{\partial^2 t_f}{\partial\rho_1\partial\rho_{2}} &
\frac{\partial^2 t_f}{\partial\rho_{2}^2}
\end{array}
\right]^{-1}
\left[
\begin{array}{c}
\frac{\partial t_f}{\partial\rho_1}\\
\frac{\partial t_f}{\partial\rho_{2}}
\end{array}
\right]
\] 
The calculations were performed using an interpolation formula,
\begin{equation}
\Phi^{RSA}(\rho) = (1-x)^3(1+b_1 x +b_2 x^2 + b_3 x^3)
\end{equation}
where $x=\rho/\rho_{\infty}$ and  $b_1$,  $b_2$, and $\rho_{\infty}$ are  taken from
reference  \cite{schaaf89}.   Fig. \ref{fig:twostep}  shows $\rho_1$ and
$\rho_2$   which   minimize the time required    to  produce an adsorbed
configuration with a  final density of $\rho_f$.  For  a given value  of
$k_d$,  the two curves converge  at  $\rho_{f0}$.  If  the desired final
density is larger  than this  value, it  will be more  efficient i.e.,
require less time, with  a two-step process   than with a  single step
process. If the final density is less  than $\rho_{f0}$, the single step
process is more efficient. As   $k_d$ increases, $\rho_{f0}$   decreases
since the time required  to remove a  given fraction of particles is a
decreasing function of $k_d$. Thus at large desorption rates less time
is spent  in the desorption phase.  We  observe that $\rho_1$ appears to
be   almost independent  of $k_d$  (for  $\rho_f>\rho_{f0}$), while $\rho_2$
curves converge at  large values of  $k_d$.   The two-step  process is
most efficient in the  limit $k_d\to\infty$. That  the two step process can
be more efficient than a one-step process  is a result of the inherent
memory of the adsorbed configurations. Without this memory effect, the
two-step process is always less efficient.

\section{Conclusion}

We have  developed a  theoretical approach  for modeling a  three-step
surface filling process composed  of two irreversible adsorption steps
separated by a desorption step. This approach considers the effects of
surface   density   and coverage  history  on   the adsorbate particle
distribution functions and predicts  an enhancement of  the saturation
surface density that is maximized when one third  of the molecules are
removed during the intermediate desorption step.

One  experimental realization of  this process is partially reversible
macromolecular adsorption where only a  certain fraction of  molecules
desorb during  a buffer  rinse.  Partial reversibility may be   due to
adsorbent heterogeneity, formation of  stable clusters of molecules on
the surface,  or  post-adsorption   transitions  in  conformation   or
orientation.   Another experimental realization of  this proces is the
elution   of  otherwise  irreversibly  adsorbed  macromolecules   with
surfactant or detergent eluting agents \cite{wahlgren98}.  The results
presented  here could have important  implications for these and other
multi-step filling processes  when a high  saturation density and/or a
rapid surface filling is required.

\section{Acknowledgments}

P. R. Van Tassel thankfully acknowledges a NATO-NSF Postdoctoral Fellowship.
J. Talbot and P. R. Van Tassel thank the NSF for financial support.
The Laboratoire de Physique Th{\'e}orique des Liquides is a Unit{\'e} Mixte de
Recherche no. 7600 associated with the Centre National de la Recherche 
Scientifique.

\newpage

\begin{figure}

\label{fig:avsf}
\caption{
The available surface function plotted as a function of density
at various fixed densities prior to and following desorption. The 
bottom line is properly placed and each subsequent line is shown
displaced vertically by 0.1 units for clarity.}
\end{figure}

\begin{figure}

\label{fig:jlden}
\caption{
The jamming limit density, as determined from the interpolation
formula (\ref{interpolant}), as a function of (a) the density, $\rho_1$
at which desorption begins and (b) the density, $\rho_2$, at
which desorption ends.
}
\end{figure}

\begin{figure}
\label{fig:sim1d}
\caption{
Enhancement of the jamming limit coverage, $\rho_f$,
in one dimension resulting from a random desorption from
a jamming RSA configuration of hard rods on a line. Each
particle is desorbed from this configuration with probability
$p_{des}$ before re-adsorption to a new jamming limit.}
\end{figure}

\begin{figure}

\label{fig:sim2d}
\caption{
Enhancement of the jamming limit coverage, $\rho_f$,
in two dimensions
resulting from a random desorption at a density $\rho_1$ to
a density $\rho_2$ as determined in a numerical simulation
of the process. $+$: $\rho_1=0.53$, $*$: $\rho_1=0.5$,
diamonds: $\rho_1 = 0.4$.}
\end{figure}

\begin{figure}

\label{fig:twostep}
\caption{Kinetics of an adsorption-desorption-adsorption process.
(a) For a given desorption rate, $k_d$, the values of the densities $\rho_1$
and $\rho_2$ that minimize the time required to reach the final coverage
$\rho_f$.(b) Comparison of the optimized two step process with $k_d=1$ and
optimized for $\rho_f = 0.5$ with
an uninterrupted adsorption process. Note that the final coverage is 
attained more rapidly using the two-step process.}
\end{figure}

\newpage
\begin{table}
\caption[99]{
Enhancement of the Saturation Coverage. 
Particles are desorbed randomly from an RSA configuration at 
an initial density $\rho_1$ until the density reaches $\rho_2$.
A new RSA process then begins on the depleted configuration. The
density at saturation is then estimated by extrapolating the
coverage versus time curves using the asymptotic power
law, (\ref{asymp}). Each condition used 800 independent averages
for a system with a relative area of the particle to simulation
cell of 0.001. For $\theta_1=0.53$ a total of $10^7$ attempts 
were made to add new particles, while for $\theta_1=0.50$ and
$\theta_1=0.4$ we used $5 \times 10^6$ attempts.}
\begin{center}
\begin{tabular}{|ccc|}\hline
$\rho_1$ & $\rho_2$ & $ \rho_{\infty}$ \\ \hline
0.53 & 0.05 & 0.5474 \\
& 0.10 & 0.5486 \\
& 0.15 & 0.5500 \\
& 0.20 & 0.5511 \\
& 0.25 & 0.5524 \\
& 0.30 & 0.5539 \\
& 0.35 & 0.5546 \\
& 0.40 & 0.5544 \\
& 0.45 & 0.5529 \\
& 0.50 & 0.5497 \\
\hline
0.50 & 0.05 & 0.5473 \\
& 0.1 & 0.5477 \\
& 0.15 & 0.5484 \\
& 0.20 & 0.5496 \\
& 0.25 & 0.5499 \\
& 0.30 & 0.5504 \\
& 0.35 & 0.5504 \\
& 0.40 & 0.5503 \\
& 0.45 & 0.5491 \\
\hline
0.45 & 0.05 & 0.5474 \\
& 0.10 & 0.5472 \\
& 0.15 & 0.5477 \\
& 0.20 & 0.5478 \\
& 0.25 & 0.5479 \\
& 0.30 & 0.5482 \\
& 0.35 & 0.5476 \\
& 0.40 & 0.5473 \\
\hline
\end{tabular}
\end{center}
\label{tab:spheres}

\end{table}


\newpage

\begin{thebibliography}{99}
\bibitem{andrade86}J. D. Andrade and V. Hlady, Adv. Polymer Sci., {\bf 86}, 1 
(1986)
\bibitem{wahlgren95}M. Wahlgren, T. Arnebrant and I. Lundstr{\"o}m,
J. Coll. Int. Sci., {\bf 175} 506, (1995)
\bibitem{buijs96}J. Buijs, P. A. W. Van den Berg, J. W. Th. Lichtenbelt,
W. Norde and J. Lyklema, J. Coll. Int. Sci., {\bf 178}, 594 (1996)
\bibitem{ramsden93a}J. J. Ramsden, J. Stat. Phys., {\bf 73}, 853 (1993)
\bibitem{robeson96}J. L. Robeson and R. D. Tilton, Langmuir, {\bf 12},
6104 (1996)
\bibitem{adamczyk94}Z. Adamczyk, B. Siwek, M. Zembala and P. Belouschek,
Adv. Colloid Interface Sci. {\bf 48}, 151 (1994)
\bibitem{cullen94}D. C. Cullen and C. R. Lowe, J. Colloid Interface Sci.,
{\bf 166}, 102 (1994)
\bibitem{evans93}J. W. Evans, Rev. Mod. Phys., {\bf 65}, 1281 (1993)
\bibitem{ramsden93}J. J. Ramsden, Phys. Rev. Lett., {\bf 71}, 295 (1993)
\bibitem{schaaf89}P. Schaaf and J. Talbot, Phys. Rev. Lett. 
{\bf 62}, 175 (1989); J. Chem. Phys., {\bf 91}, 4401 (1989)
\bibitem{tarjus91}G. Tarjus, P. Schaaf and J. Talbot, J. Stat.
Phys., {\bf 63}, 167 (1991)
\bibitem{jin94}X. Jin, Ph.D. thesis, Purdue University (1994)
\bibitem{ricci92}S. M. Ricci, J. Talbot, G. Tarjus and P. Viot,
J. Chem. Phys., {\bf 97}, 5219 (1992)
\bibitem{adamczyk96}Z. Adamczyk, B. Siwek, M. Zembala and P. Weronski,
J. Chem. Phys., {\bf 105}, 5562 (1996)
\bibitem{johnson95}P. R. Johnson and M. Elimelech, 
Langmuir, {\bf 11}, 801 (1995)
\bibitem{vantassel94}P. R. Van Tassel, P. Viot, G. Tarjus and J. Talbot
J. Chem. Phys., {\bf 91}, 7064 (1994)
\bibitem{vantassel96}P. R. Van Tassel, J. Talbot, G. Tarjus
and P. Viot, Phys. Rev. E, {\bf 53}, 785 (1996)
\bibitem{kurrat94}R. Kurrat, J.J. Ramsden and J.E. Prenosil, J. Chem. Soc.
{\bf 90}, 587 (1994)
\bibitem{wannerberger96}K. Wannerberger and T. Arnebrant, J. Colloid
Interface Sci., {\bf 177}, 316 (1996)
\bibitem{malmsten96}M. Malmsten, B. Lassen, J. Westin, C.-G. Golander,
R. Larson and U.R. Nilsson, J. Colloid Interface Sci., {\bf 179}, 163
(1996)
\bibitem{wahlgren98}M. Wahlgren, S. Welin-Klintstr{\"o}m and C. A.-C.
Karlsson, 485, {\it Biopolymers at Interfaces}, M. Malmsten ed. Marcel Dekker,
New York (1998)
\bibitem{vantassel97}P. R. Van Tassel, J. Talbot, P. Viot and G. Tarjus
Phys. Rev. E, {\bf 56}, R1299 (1997) 
\bibitem{tilton89}R. D. Tilton, C. R. Robertson, A. P. Gast, J. Colloid Inteface Sci.,
{\bf 137}, 192 (1989)
\bibitem{norde92}W. Norde and A. C. I. Ausiem, {\bf 66}, 73 (1992)
\end{thebibliography}
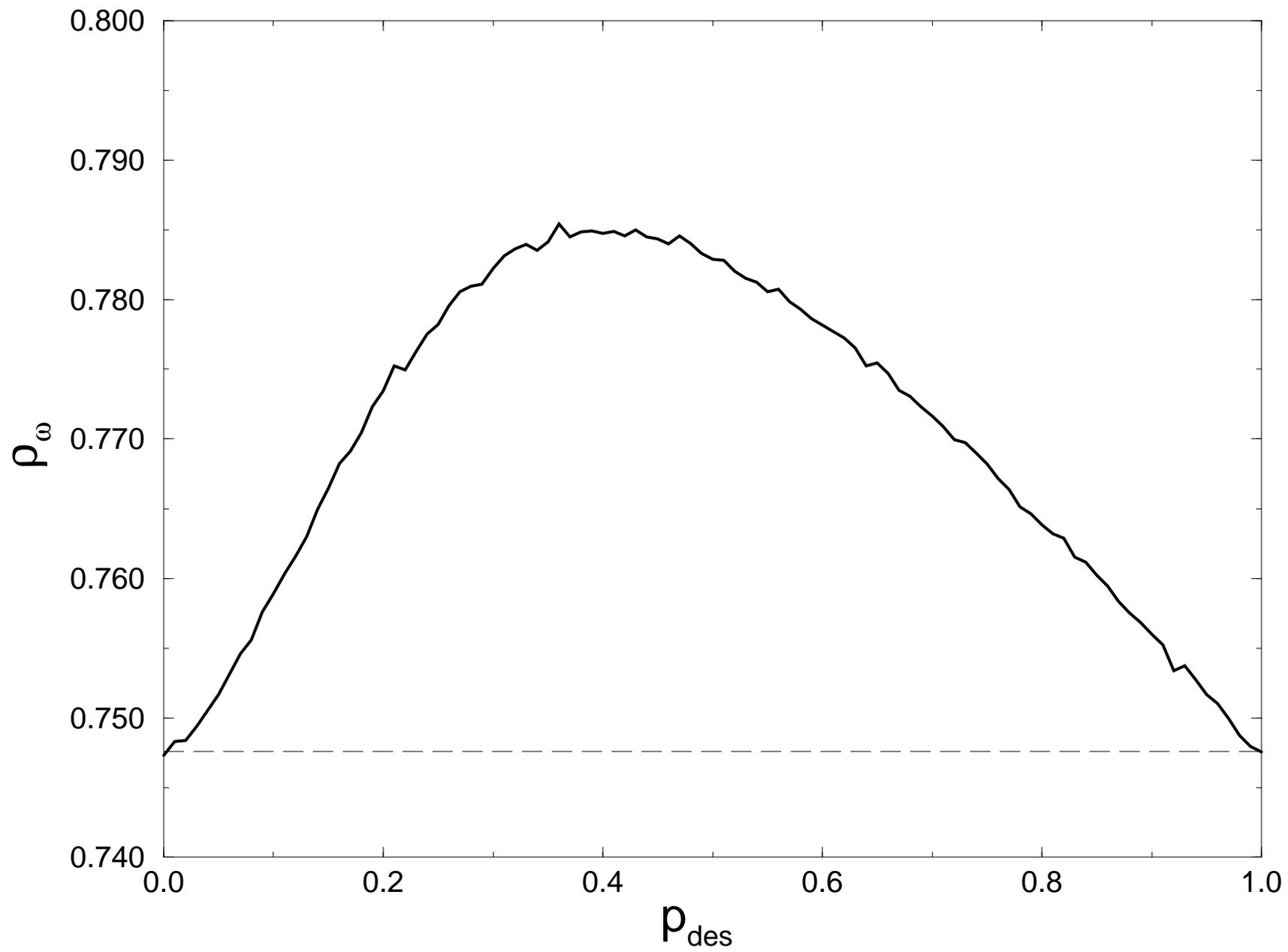
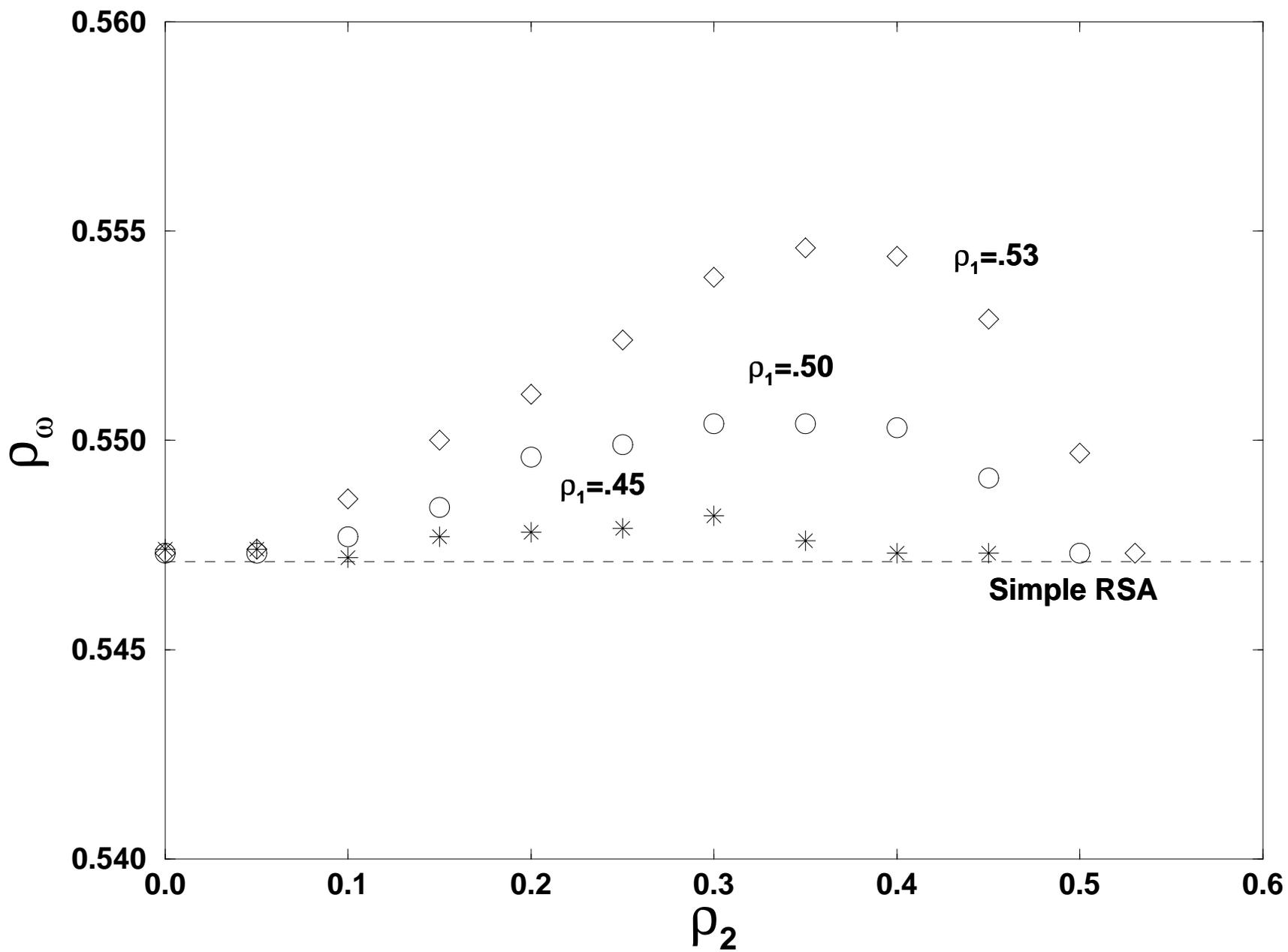
\end{document}